# System Interactive Cyber Presence for E- Learning to Break Down Learner Isolation

Bousaaid Mourad
Department of Computer Science
Faculty of Sciences, Ibn Zohr University
Agadir, Morocco

Ayaou Tarik
Department of physics
Faculty of Sciences, Ibn Zohr University
Agadir, Morocco
Afdel Karim

Department of Computer Science
Faculty of Sciences, Ibn Zohr University
Agadir, Morocco

Estraillier Pascal
L3i Laboratory
University of La Rochelle
La Rochelle, France

## ABSTRACT

The development of technologies of multimedia, linked to that of Internet and democratization of high speed, has made henceforth E-learning possible for learners being in virtual classes and geographically distributed. One benefit to taking course online is that the online course structure is typically more student focused than teacher centered and encouraging more active participation by students in collaborative learning activities. The quality and quantity of asynchronous and synchronous communications are the key elements for E-learning success. A potential problem that has received little exploration is student's feeling of isolation. It is important to have a propitious supervision to breaking down learner feeling isolation in E-learning environment. This feeling of isolation is among the main causes of loss and high rates of dropout in E-learning. It impacts on their levels of participation, satisfaction and learning. To overcome this feeling of isolation, we aim, by this research, to provide the trainer and each learner with an environment allowing them to behave as if being face to face; in other words, to approach the pedagogy of classroom teaching. Our contribution to reduce the feeling of isolation is to ensure the presence of the teacher in the educational tools. These tools aim to establish a real dialogue with the learner, forcing him to take an active part in their learning. Among the tools we offer, video conference Openmeeting integrated in Moodle providing the possibility of using the notion of class and whiteboard, the indicator of motivation quantification tool based hand gesture that we developed and finally social networks web 2.0 like Facebook, youtube, twitter… to promote collaboration, sharing and communication of the learner with his peers.

## 1. INTRODUCTION

Traditionally, E-learning implies an individualized asynchronous and/or synchronous learning. The pedagogy of E-learning relies mainly on the learner's management of learning process. This pedagogy has a direct impact on the teaching materials, the offered educational support and the realized learning. In a distance course, the teaching materials should be the same as a teacher would use in real class. Therefore, they are not a simple course text. The preparation of these teaching materials is based on a pedagogical approach which involves many tools where interactivity plays a role of great importance. During distance learning, even the most motivated and organized learner may be at risk of abandoning along the way, due to the isolation that might be felt. In order to maximize the continuity of the teacher presence in teaching materials[1,2] should be guaranteed by including in the course content a real dialogue with the learner, forcing him to take an active part in their learning. For this end, the use of new technologies, either in full distance learning or classroom supported by ICT, is extremely important.

Thanks to ICT technologies, there might be some convergence between the pedagogy of classroom teaching and pedagogy of distance learning. The first pedagogy could benefit from the basic principles of distance learning to make the teacher get rid of his role as "information transmitter" and enable him to offer more individualized instruction by acting as guide and adviser using, for instance, the pedagogy of the inverted class. The second pedagogy could enrich the educational support offered to students and allow them to communicate, so as to reduce the inherent isolation feeling in distance learning and reconnect with the pleasure and motivation provided by social interaction in the classroom. It is therefore essential to provide a suitable supervision to reduce the feeling of isolation and passivity which is one of direct causes of loss and high dropout rates in distance learning. To overcome this feeling of isolation, we aim, by this research, to provide the trainer and each learner with an environment allowing them to behave as if being face to face; that is, to opt for the approach of classroom teaching pedagogy. One of the preferred means is the use of USB cameras to analyze in real-time movements related to the behavior of the trainer and learners especially consideration of hand gestures as an interactive technique that can potentially provide more natural, intuitive and creative methods to communicate and quantify the learner's attention and his level of participation. Also, it will allow the trainer to have a "feedback" quality of learners' reactions similar to those he is accustomed to in classroom by using tools as openmeeting integrated in Moodle providing the possibility of using the notion of class and whiteboard and social networks like Facebook to promote collaboration, sharing and communication of the learner with his peers. In this paper, we will define E-learning platform in Section 2. The Section 3 will be devoted to the hand gesture tracking tool. This gesture will be analyzed and transformed into a participation indicator. In Section 4, we use web 2.0 "toolbox" to improve communication, collaboration and sharing among students and teachers before summing up the outcomes of the paper in the conclusion.

## 2. E-LEARNING PLATFORM

### 2.1 Environment

We will replace the absence of the teacher and the social interaction within the classroom through pedagogical tools to reduce the feeling of isolation. Our system consists essentially of the combination between the learning environment Moodle[3,4] and the videoconferencing solution Open Meetings[5]. Moodle is a free open source Course Management System (CMS), also known as a Learning Management System (LMS) [3 used by more than 30 000 educational organizations around the world [2]. Open meetings, embedded into Red5 [3], "allows to set up instantly a conference in the Web and to use microphone or webcam, share documents on a white board, share screen or record meetings". The aim of using this tool in the platform is to complement Moodle on the synchronous communications activities. Our system meets the features of the presential courses according to the analysis done in the first chapter.

### 2.2 In the Class Room

We will now provide use case diagrams and significant sequence of our environment (Fig. 1). This diagram shows the process of class attendance that is accomplished by the connection of the actors (teachers, students). The diagram explains the mechanism of audiovisual communication between members of the class. The environment provided by the Open Meetings component consists essentially of an audio video stream manager, a whiteboard and a chat area. The teacher plays the role of moderator.

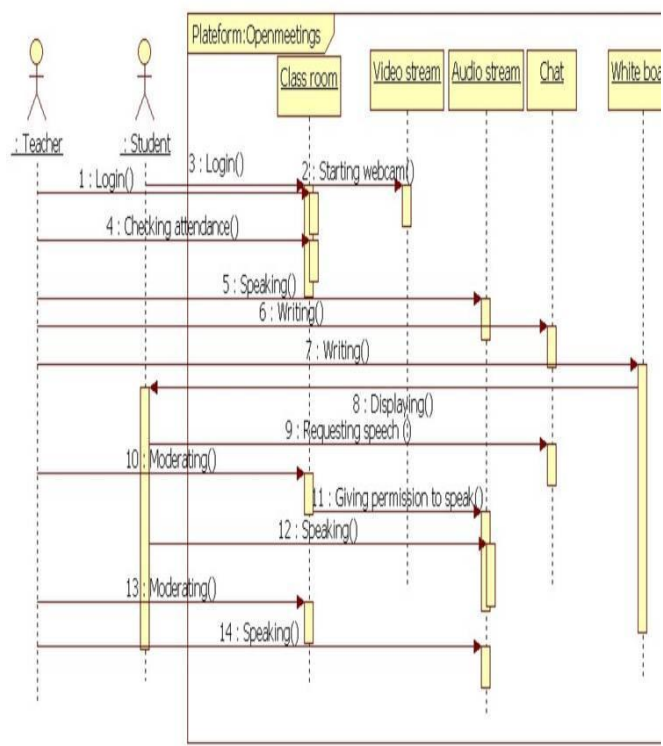

**Fig 1: Implementation of the detection system, the recognition of hand gesture**

### 2.3 On the White Board

The face to face learning is characterized by the ability to use the white board by [5] a learner following the teacher's request. This is now possible thanks to the whiteboard managed by the moderator in Openmeeting as indicated by the following scheme (Fig.2).

The course animator plays the role of moderator in Openmeeting with the opportunity for the student to use the board by his initiative or at the request of the teacher.

### 2.4 Distant Demonstration

To resolve an eventual problem of learner's dropout because of technical problems, the platform contains integrated consoles which enable the moderator to share the screen in order to assist the students. The following diagram (Fig. 2) describes this mechanism:

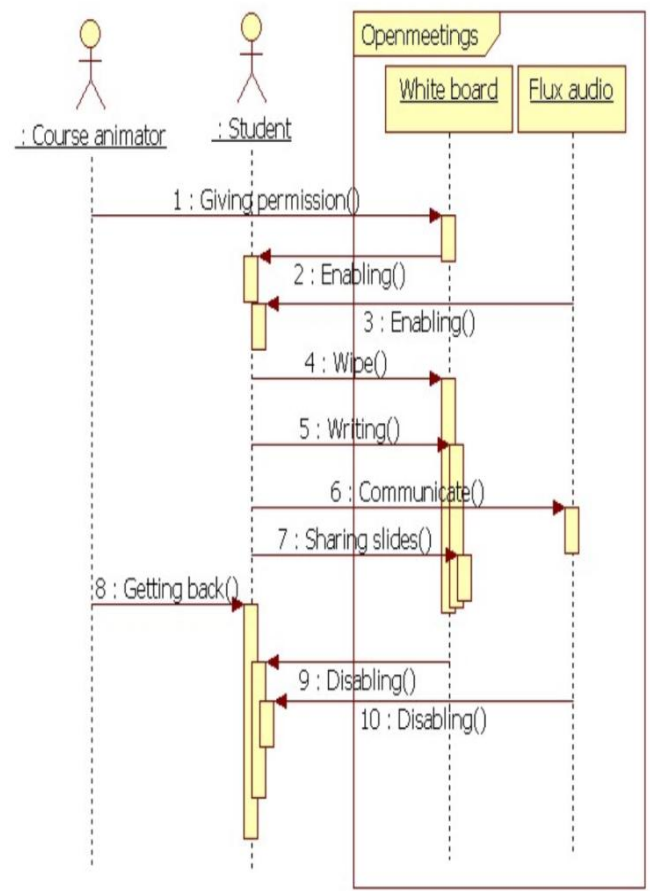

**Fig. 2: Sequence diagram of distant demonstration**

## 3. Indicator of Motivation Quantification Tool based on Hand Gesture

Hand gesture is one of the most important tools for communication in human daily life. For instance, while learning, such gesture can be used to communicate or stop the tutor if learner does not understand some concepts, or he wants to answer questions intermittently posed by the tutor. These gestures can be quantified as an indicator of learner attention and his level of participation during a learning session. We have developed a tool for quantifying the participation indicator in [25]. The summary of this research work is presented below.

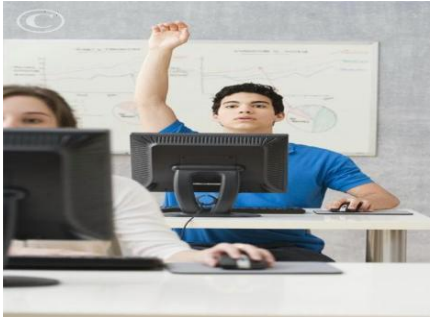

**Figure 3: Use of hand gesture as an indicator of participation**

This system is composed of following stages:

- Stage of processing and detection of hand gesture;
- Stage of hand gesture recognition

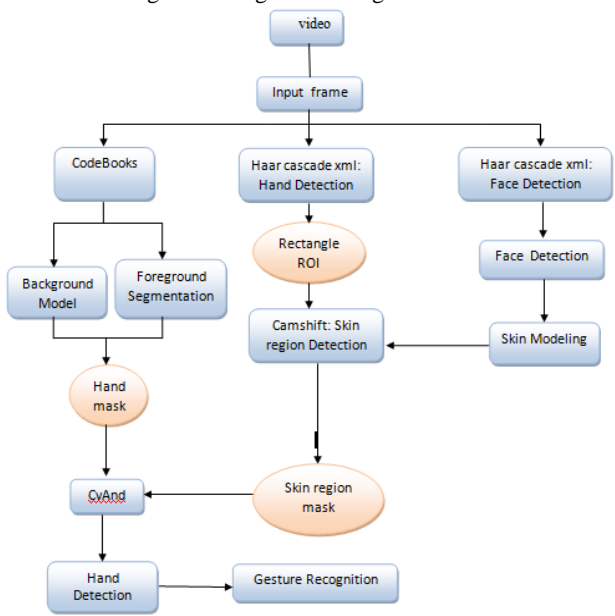

**Fig. 4: Implementation of the detection system, the recognition of hand gesture**

## 3.1 Hand Gesture Detection

To be able to detect hand gesture, there are many processes to be carried out so as to eliminate information which may damage this stage of detection like light variation and problems of static and dynamic video sequences background:

### 3.1.1 Elimination of Noise Video Sequences Background

In this stage, we eliminate useless information of compound video sequences background and problems of light changes.

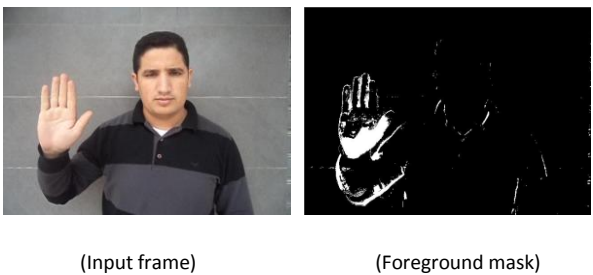

(Input frame)                    (Foreground mask)

**Figure 5: examples of first binary mask of moving hand gesture with codebook**

### 3.1.2 Detection of the Hand based on Haar Cascade Method:

Haar Cascade is a method of image object detection, proposed by searchers Paul Viola and Michael Jones [15]. It is a part of all first methods capable of detecting effectively and in real time image objects. Originally invented to detect faces [15], it may also be used to detect other types of objects like cars or planes. In our work, we have relied on an existed haar Cascade to detect the hand.

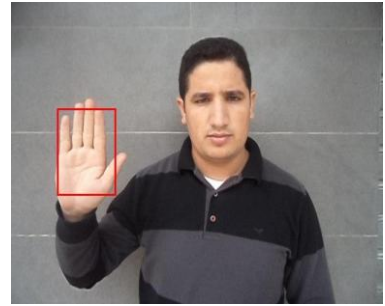

**Figure 6: hand detection using our Haar cascade xml: Hand**

Once the hand gesture is detected, we can provide coordinates of rectangle surrounding hand gesture or what we call ROI, which will be the next initial stage that will locate hand position for CamShift function.

### 3.1.3 Detection of Skin by CamShift Technique

To detect the skin, we use Camshift method. This technique uses the skin color to construct the second mask of hand gesture movement. The problem of this technique is its sensitivity to light variation. To solve this problem and facilitate skin detection, a part of the face is chosen using Haar Cascade to detect the face of Jones and Viola [10]. Our purpose is selecting the skin pixels, then calculating statistics on adaptable distribution of skin color and defining also the model representing the skin color [13][16]. After thresholding and adequate morphological processing, we can obtain from this model a second mask of hand gesture movement. We will combine the two masks obtained on the same frames using the "and" operator to remove the pixels corresponding to the face and keep only the pixels belonging to the hand.

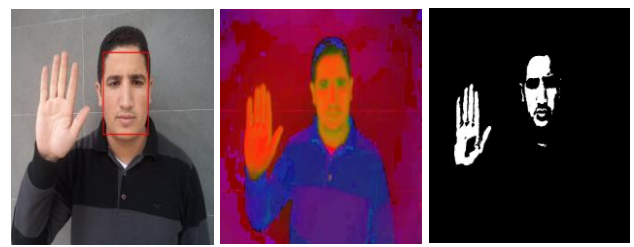

**Fig. 7: a: face detection with harr cascade face, (b): HSV model, (c): Skin region Detection mask after thresholding and adequate morphological processing**

## 3.2 Recognition of Hand Gestures with Shape Feature Contour Point Distribution Histogram (CPDH)

Once we got the mask corresponding to the hand gesture, we will move to the stage of gesture recognition. For this, we use shape feature CPDH technique [9]. This technique is based on the distribution of points on object contour under polar-coordinates. In this algorithm, the object boundary is detected using standard Canny operator to describe its shape. The result points on the contour can be represented as, $P = \{(x_1, y_1), (x_2, y_2), ..., (x_n, y_n)\}, (x_i, y_i) \in R^2$, where n represents the total number of points on contour. After the extraction of points on object boundary, the centroid of the hand boundary is computed. The maximum value of distance between the centroid and the boundary points on the contour is chosen as the radius of the minimum circumscribed circle.

The following figure shows the algorithm for the construction of CPDH [9] :

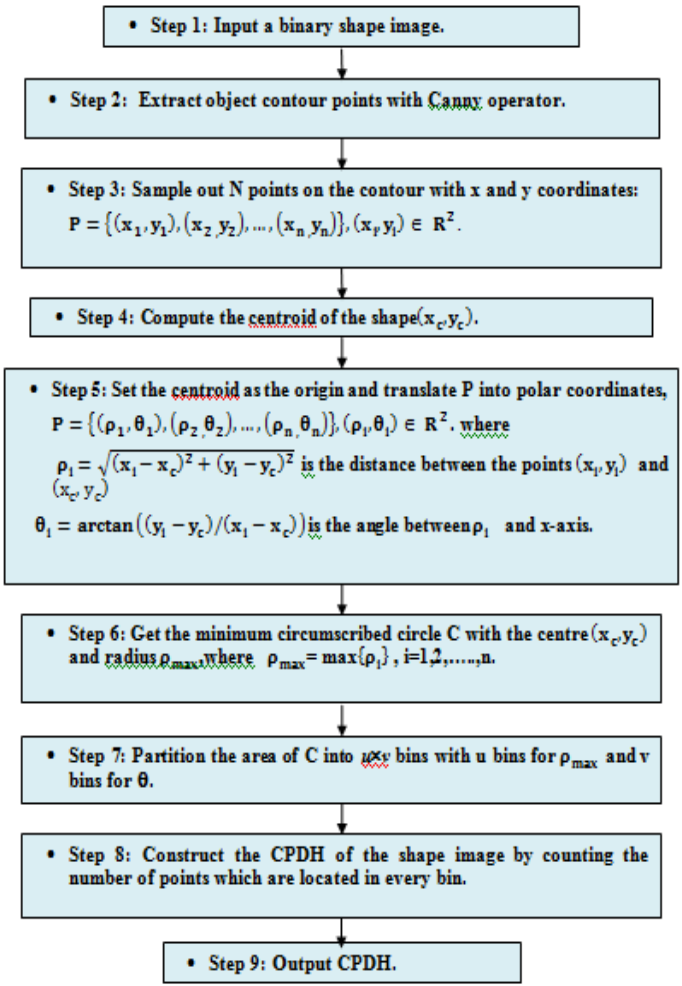

**Fig. 9: algorithm for the construction of CPDH**

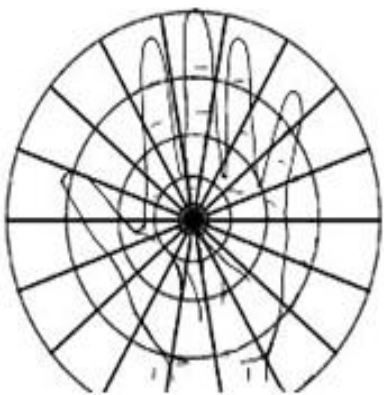

**Figure 10: CPDH shape feature**

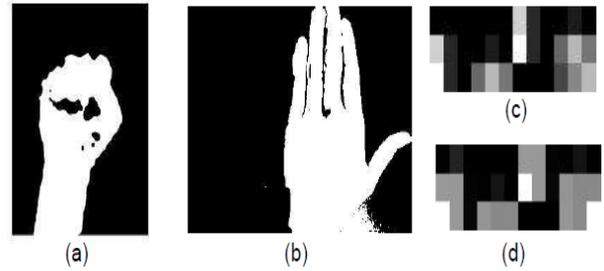

**Fig. 11: (a) and (c): binary mask hand, (b) and (d): a CPDH's constructor correspondent.**

## 3.3 Distributed Module to Supervise Learner's Participation

The fact that learners belong to different geographic sites, and that many of them follow the same synchronous learning session, it would be difficult to observe, by video direct use, the level of participation of every learner. For this, we are opting for a system based on COBRA distributed technology to qualify the attention and level of participation of learners. This system will solve the problem of heterogeneity of platforms and programming languages (C++ and Java) used in this work.

Distributed application is to be composed of modules that should be installed either beside the learner or tutor. Every learner should have a high connection output and an active USB camera. The first sub-system, installed at learner's side, allows detection, recognition and tracking of the hand gesture of the learner, while answering questions of the tutor or simply asking for help. The second sub-system, installed at tutor's side, will control the participation and the extent of learners' attention.

Fig. 12 shows the archituctre of this sub-system of supervision

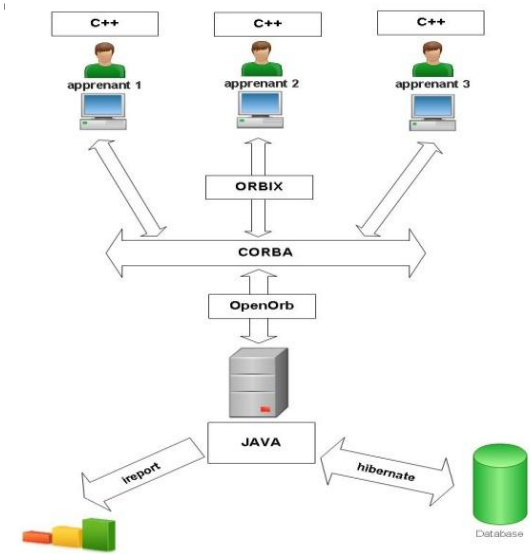

**Fig. 12 : Architecture of ditributed application based on COBRA to qualify the participation indicator of learners.**

Figure 13 and 14 show an example of participation curve of four students during a synchronous learning session. The participation of the learner 1 is very active which is translated as an important value of participation indicator, whereas such indicator has a very low value for the learner 4 who is certainly facing difficulties and problems the tutor should figure out.

activity [20]. The Web 2.0 stands out from the Web 1.0 by the

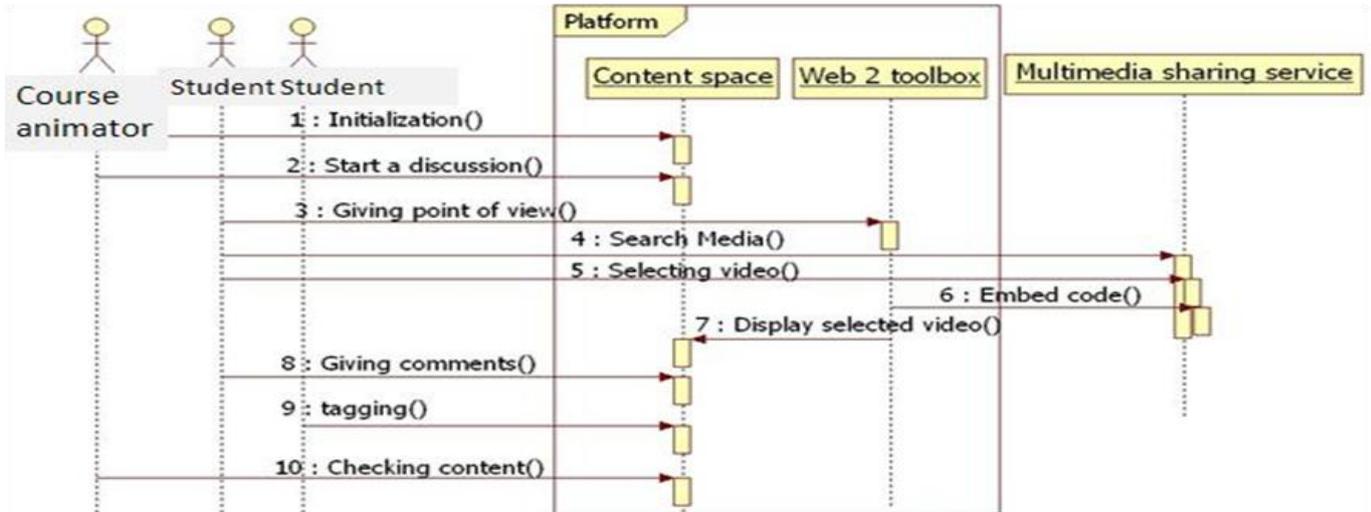

**Fig. 15: Sequence diagram of web 2.0 "toolbox" use**

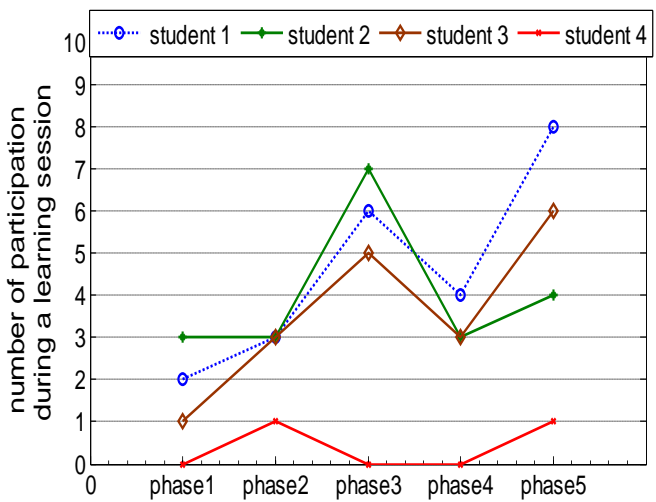

**Fig. 13: participation curve of learners during a session learning.**

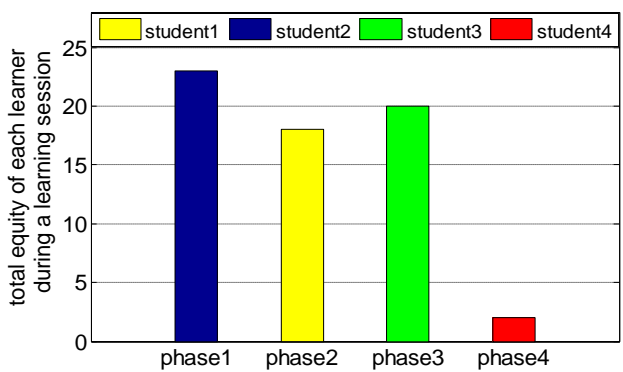

**Fig. 14: quantification of learners' participation indicator during session learning.**

## 4. WEB 2.0 "TOOLBOX" EMBEDDED AND USE

The web has become the most significant technology of the 21st century and has deeply impacted every sphere of our ease of publishing, collaborating, and sharing.

Since Web 2.0 features are gaining an immense interest by learners, it is useful to use this trend by feeding educational discussion through sharing, commenting and tagging [21]. Thus, e-Learning is playing an important role to life-long learning [22]. To achieve this mission, the platform integrates a component called "Web2 toolbox" which is a powerful tool for education. It allows users (teacher and students) to get pedagogical resources from popular services which facilitate hosting, downloading and sharing over the internet like YouTube, Socialgo, Flicker and Twitter. It involves them in content creation and enhances their education. It also improves communication among students and teachers through forums and blogs [23].

The Web2 toolbox aims to convey users' messages simply so that it respects rules and the code of practice.

Content can be a forum, a blog, a wiki [18] or micro blogging [19]. The teacher starts the discussion. Through the participation of different actors (learners and teachers) the content is enriched. They can use multimedia documents to tag and to comment. YouTube, Flicker and Twitter are examples for Multimedia sharing service websites in fig. 10.

## 5. CONCLUSION

The objective of this research is to provide the trainer and every learner with an environment that allows them to behave as if being in real class. One of the preferred ways to break down learner isolation is the use of USB cameras to analyze in real-time movements related to the behavior of the trainer and learners especially consideration of hand gestures as an interactive technique that can potentially provide more natural, intuitive and creative methods to communicate and quantify the learner's attention and his level of participation. Also, it will allow the trainer to have a "feedback" quality of learners' reactions similar to those he is accustomed to in real classroom by using tools as openmeeting integrated in Moodle. The latter provides the possibility of using the notion of class and whiteboard and social networks like Facebook to promote collaboration, sharing and communication of the learner with his peers.